\documentclass[prb,aps,twocolumn,showpacs,superscriptaddress,amsmath,amssymb,floatfix,10pt]{revtex4}
\usepackage{graphicx}
\usepackage{epstopdf}

\newcommand{\be}{\begin{equation}}
\newcommand{\ee}{\end{equation}}

\newcommand{\Tr}{{\rm Tr}}

\newcommand{\eq}[1]{(\ref{eq:#1})}

\newcommand{\ddt}{\frac{\partial}{\partial t}}

\begin{document}
\title{Wave function Monte Carlo method for polariton condensates}
\author{Michiel Wouters}
\affiliation{TQC, Universiteit Antwerpen, B-2020 Antwerpen, Belgium.}
\begin{abstract}
We present a quantum jump approach to describe coupled quantum and classical systems in the context of Bose-Einstein condensation in the solid state. In our formalism, the excitonic gain medium is described by classical rate equations, while the polariton modes are described fully quantum mechanically. We show the equivalence of our method with a master equation approach. As an application, we compute the linewidth of a single mode polariton condensate. Both the line broadening due to the interactions between polaritons and the interactions with the reservoir excitons is taken into account.
\end{abstract}
\maketitle

\section{Introduction}
\begin{figure}
	\includegraphics[width=0.5\columnwidth]{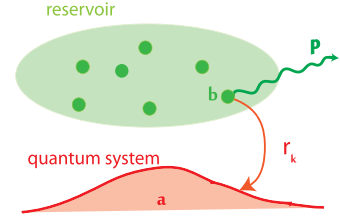}
	\caption{Sketch of the problem under study: The incoherent reservoir particles relax at a rate $r_k$ to the quantum sytem by giving the excess energy to the environment modes $p$.}
	\label{fig:sketch}
\end{figure}

The achievement of strong light-matter coupling in semiconductor micro~\cite{deveaud_book} and nanocavities \cite{hennessy,senellart} has led to the manifestation of several novel quantum effects in the solid state. A prominent example is Bose-Einstein condensation of exciton-polaritons \cite{kasprzak}. A second important topic is the dynamics of nanocavities in the strong coupling regime with a quantum dot, where lasing was recently observed~\cite{arakawa}. These phenomena are characterized by a large degree of complexity, due to the coupling of quantum dynamics with the nontrivial dynamics of the injected incoherent particles. 
The physics that these solid state systems have in common is that there are some degrees of freedom for which the quantum mechanical nature is important, while other degrees of freedom are essentially classical. This is illustrated in Fig.~\ref{fig:sketch}. Due to their coupling, it is not straightforward to construct a model for the combined dynamics of the classical and quantum parts. The typical approach is an adiabatic elimination of the classical reservoir, assuming that it adapts instantaneously to the state of the quantum part~\cite{imamoglu-yamamoto}. For polariton condensation, this type of approach was developed by Laussy et al. \cite{laussy}. Also in Keldysh Green function treatments, the bath degrees of freedom are usually integrated out \cite{keeling}, which again limits the types of interactions between bath and system that can be taken into account.

In this Article, we want to construct a theory that goes beyond such approximations and does not make any other approximations apart from a classical treatment of the reservoir and that the coupling between the reservoir and quantum system is weak. We will show that a master equation approach leads in general to an infinite hierachy of equations, that does not allow for an easy truncation. We subsequently show that a quantum jump model can be constructed that is equivalent to the infinite set of coupled master equations. We explicitly show that the Schawlow-Townes linewidth and Henry linewidth enhancement factor~\cite{henry,haug_book} and line broadening due to polariton-polariton interactions \cite{eastham} are reproduced with our model. 

\section{Master Equation}

We will derive our equations for the quantum evolution of the coupled polariton-reservoir system, assuming the interaction between the two to be of the form
\begin{equation}
H_I=\sum_k g_{kq} p^\dag_q b^\dag_k a + {\rm h.c.}
\label{eq:HI}
\end{equation}
This Hamiltonian describes the relaxation process illustrated in Fig.~\ref{fig:sketch}: a reservoir particle $b_k$ relaxes to the quantum system mode $a$, depositing its excess energy in the `phonon' mode $p_q$, that we assume for simplicity to be at zero temperature. In the usual Born-Markov approximation, the following master equation can be derived for the reduced density matrix
\begin{equation}
\frac{\partial}{\partial t} \Tr_R(\rho)= \sum_k  \frac{r_k}{2}
\Tr_R \left( a^\dag b_k \rho b^\dag_k a- b^\dag_k b_k a a^\dag \rho
+ {\rm h.c.}
\right)
\label{eq:master1}
\end{equation}
under the assumption that coherence in the reservoir is negligible, so that e.g. $\Tr_R[b^2 (a^\dag)^2 \rho]=0$. The rate $r_k$ is the Fermi golden rule transition rate from the state $b_k$ into states $a$ and $p_q$.

Unless the reservoir density is assumed to be not affected by the polariton dynamics, Eq.~\eq{master1} is not a closed equation of motion for the polariton density matrix $\rho_S=\Tr_R(\rho)$, but it is coupled to the reservoir weighted density matrices $\rho_k = \Tr_R(b^\dag_k b_k\rho)$ as
\begin{equation}
\frac{\partial}{\partial t} \Tr_R(\rho)= \sum_k  \frac{r_k}{2}
\Tr_R \left( a^\dag \rho_k a- a a^\dag \rho_k
+ {\rm h.c.}
\right)
\label{eq:master1b}
\end{equation}

The dynamics of $\rho_{k}$ is under the same assumptions that lead to Eq.~\eq{master1} described by
\begin{equation}
\frac{\partial}{\partial t} \rho_k = - \sum_j \frac{r_j}{2}
\left( a a^\dag \rho_{j,k} + \rho_{j,k} a a^\dag 
\right)
+ {\mathcal P_j},
\label{eq:master2}
\end{equation}
where we have introduced $\rho_{k,j}=\Tr_R( b^\dag_k b_k b^\dag_j b_j \rho)$. 

For the pumping term $\mathcal P$ in Eq.~\eq{master2}, we take a standard gain mechanism that does not include gain saturation (the physical pump is assumed to be for example an external laser that excites the reservoir), leading to
\begin{eqnarray}
{\mathcal P_j} &=& \Tr_R\left[
b^\dag_j b_j \frac{P}{2} \left(
b_j^\dag \rho b_j + b_j b^\dag_j \rho +{\rm h.c.}
\right)
\right] \\
&\approx&  P \Tr_R\left( \rho \right).
\label{eq:pump}
\end{eqnarray}
For the last step, we have assumed that the occupation of each mode is on average much lower than one, so that we can neglect $\Tr(b^\dag_k b_k \rho)$ whith respect to $\Tr(\rho)$. 

\section{Quantum jump model}

It is clear that the dynamics of $\rho_{j,k}$ will involve the expectation value of densities in three reservoir modes and so on. 
Physically, Eq.~\eq{master1} expresses that gain is proportional to the density matrix weighted with the reservoir densities. For weak gain saturation, i.e. when the number of reservoir particles weakly depends on the polariton number, the gain saturation can be treated perturbatively~\cite{imamoglu-yamamoto}. It is the aim of this article to go beyond such a perturbative treatment and propose a wave function Monte Carlo method (see e.g. \cite{breuer}) that is equivalent to the hierarchy of Eqns.~\eq{master1},~\eq{master2}, etc. 

Because the reservoir is assumed to be classical, its state can be described by the occupations alone $r=\{n_1,n_2,\ldots\}$. The polariton part is treated fully quantum mechanically by a wave function $| \Phi \rangle$. The state $S$ of the system and reservoir is then $S=[r,|\Phi \rangle ]$. A stochastic dynamics for $S$ leads to a probability distribution $P(S)$, that has an associated density matrix
\begin{equation}
\rho = \int d \Phi \sum_{r_1,r_2,\ldots} P(\{n_i\},|\Phi \rangle)  |\Phi \rangle \langle \Phi |,
\label{eq:rho}
\end{equation}
where the integration runs over all possible values of the wave function.
The reservoir weighted density matrices are defined analogously as
\begin{equation}
\rho_k = \int d \Phi \sum_{r_1,r_2,\ldots} P(\{n_i\},|\Phi \rangle)\; n_k |\Phi \rangle \langle \Phi |.
\label{eq:rhok}
\end{equation}

For the deterministic part of the dynamics, we propose the nonlinear dynamics that is standard in wave function Monte Carlo: the reservoir densities are invariant under the deterministic part of the evolution $dn_i/dt=0$, where the wave function evolves according to 
\begin{equation}
i\frac{\partial }{\partial t} |\Phi \rangle = 
\frac{-i}{2} \sum_k r_k n_k a a^\dag |\Phi \rangle + \frac{i}{2} R(\{n_i\})
 \|  a^\dag  |\Phi \rangle \|^2 |\Phi \rangle,
\label{eq:deterministic}
\end{equation}
where $R(\{n_i\})=\sum_k r_k n_k$ is the total spontaneous rate for the creation of polaritons out of the reservoir. The second term in Eq. \eq{deterministic} compensates for the norm-reducing dynamics due to the first term.
After an evolution over time $\delta t$, the wave function is transformed to
\begin{equation}
|\Phi(t+\delta t)\rangle = \frac{1}{\sqrt{1-\delta p}}\left[|\Phi(t)\rangle 
-  \frac{R (\{n_i\})\delta t}{2}   a a^\dag |\Phi \rangle \right],
\label{eq:dpsi}
\end{equation}
where 
\begin{equation}
\delta p=\delta t \,R({n_i}) \|  a^\dag  |\Phi \rangle \|^2
\label{eq:deltap}
\end{equation}
 is the probability that a polariton is created out of the reservoir during the time $\delta t$.

The jump part of the dynamics due to the creation of a polariton from a reservoir particle in mode $k$ is given by
\begin{eqnarray}
n_k &\rightarrow & n_k-1, \label{eq:jump1} \\
| \Phi \rangle &\rightarrow & \frac{a^\dag | \Phi \rangle}{\|a^\dag | \Phi \rangle \|^2}.
\label{eq:jump2}
\end{eqnarray}
Such a jump occurs in an interval $dt$ with probabilities 
$\delta p_k=\delta t \,r_k n_k \|  a^\dag |\Phi \rangle \|^2$ that sum as $\sum_k \delta p_k = \delta p$.

Under the dynamics~\eq{dpsi}, \eq{jump1} and \eq{jump2}, the expectation value of the density matrix $|\Phi \rangle \langle \Phi |$ evolves as
\begin{multline}
E[|\Phi(t+\delta t)\rangle \langle \Phi(t+\delta t) |] = E [ |\Phi(t)\rangle \langle \Phi(t) | \\
-\frac{1}{2} (1-\delta p) R({n_i}) \delta t  \frac{a a^\dag | \Phi \rangle }{\sqrt{1-\delta p}} \frac{ \langle \Phi |}{\sqrt{1-\delta p}} \\
-\frac{1}{2} (1-\delta p) R({n_i}) \delta t  \frac{ | \Phi \rangle }{\sqrt{1-\delta p}} \frac{ \langle \Phi |a a^\dag}{\sqrt{1-\delta p}} \\
+ \delta p \frac{a^\dag |\Phi}{\sqrt{\delta p/\delta t}} \frac{\rangle \langle \Phi | a}{\sqrt{\delta p/\delta t}} ].
\label{eq:stoch1}
\end{multline}
With the definitions ~\eq{rhok} and \eqref{eq:deltap}, Eq. \eqref{eq:stoch1} implies that the expectation value $E[|\Phi(t+\delta t)\rangle \langle \Phi(t+\delta t) |]$ obeys the equation of motion \eq{master1}.

In the same way, one can verify that the evolution stochastic evolution for $|\Phi \rangle$ leads to the equation of motion \eq{master1b} with the definition
\begin{equation}
\rho_{j,k} = \int d \Phi \sum_{r_1,r_2,\ldots} P(\{n_i\},|\Phi \rangle)\; n_j n_k |\Phi \rangle \langle \Phi |.
\label{eq:def_rkj}
\end{equation}

The correspondence between the master equation description and the quantum jump appoach is entirely clear from a physical point of view. The classical reservoir continuously monitors the polariton mode, which leads to the quantum jumps. The present approach is a generalization of the usual quantum jump models in the sense that we take into account the backaction of the system on the reservoir, which in turn affects the system dynamics.

The pumping of the reservoir can be taken into account by the additional jump process acting on the reservoir state alone $n_k \rightarrow n_k+1$  with a probability in the interval $\delta t$ equal to $\delta p=\delta t P$. This jump process leads to a term in the evolution of the density matrix $\rho_k$ of the form
\begin{equation}
\ddt \rho_k = P \int d \Phi \sum_k r_k   |\Phi \rangle \langle \Phi |,
\label{eq:jump_pump}
\end{equation}
which reproduces Eq.~\eq{pump}. In addition, all internal dynamics in the reservoir, due to for example collisions between particles in the reservoir can be simply modeled by applying the modification on the reservoir mode occupations and keeping the polariton state the same.

The dynamics~\eq{deterministic}-\eq{jump2} naturally leads to polariton states that are in a number state, because the reservoir monitors the polariton number by the stimulated relaxation. This implies that a semiclassical stochastic model for a quantum dot embedded in a nanocavity, such as e.g. performed in Ref.~\cite{guillaume} does not make any approximations apart from the ones involved in the derivation of the quantum Markov dynamics~\eq{master1}. Moreover, this picture is consistent with the discussion of the coherence properties of laser light by M\o{}lmer~\cite{molmer}, who advocates the point of view that the photon state inside a laser cavity is a number state, rather than a coherent state.

\section{Temporal coherence}

For the computation of the temporal coherences, the stochastic wave function dynamics can be computed on a doubled Hilbert space~\cite{breuer} for the polariton state. More specifically, the first order coherence $g^{(1)}(t+\tau,t)=\langle a^\dag(t+\tau) a(t) \rangle$ can be computed by propagating the state $S_E = [r,\theta]$, where the normalized state vector $\theta(t)$ is proportional to $[\Phi(t)\;\; a\Phi(t)]^T$ over a time $\tau$ according to the prescription detailed in Ref.~\cite{breuer}, that is to apply the evolutions~\eq{deterministic} and~\eq{jump2} on the two components of the wave function with a proper normalization of the total wave vector. The temporal coherence is then given by the expectation value 
\begin{equation}
g^{(1)}(t+\tau,t)=E[\|\ \langle \Phi(t)| a |\Phi(t)\rangle \|^2 \langle \theta_2(t+s\tau) | a^\dag | \theta_1(t+\tau) \rangle]
\end{equation}
, where $\theta_1$ and $\theta_2$ refer to the first and second part of the doubled Hilbert space respectively. 

To understand the coherence properties of the polariton field, it is instructive to compute the temporal coherence of a state that is initially in a number state with a large number of polaritons $N$ that decay at a rate $\gamma$. The deterministic evolution of the state $\theta=(u |N\rangle, v|N-1\rangle)^T$ gives a change of the components on the doubled Hilbert space
\begin{equation}
\ddt
\left(
\begin{array}{l}
u |N\rangle \\
v  |N-1\rangle
\end{array}
\right)
=
\left(
\begin{array}{l}
\frac{u(t)}{u^2+e^{\gamma t}v^2} |N\rangle \\
\frac{v e^{\gamma t}}{u^2+e^{\gamma t}v^2}  |N-1\rangle
\end{array}
\right)
\label{eq:dtheta}
\end{equation}
The coherence is proportional to $u v=u \sqrt{1-u^2}$. When a polariton is lost due to a quantum jump after a time $\tau$, the change in $u$ under a deterministic evolution followed by quantum jump is
\begin{equation}
u\rightarrow \frac{\sqrt{n}u}{\sqrt{n u^2 +(N-1)(1-u^2) \exp[r \tau]}}
\label{eq:total}
\end{equation}
Under this evolution, $u$ tends to zero for long times. The temporal coherence time is thus equal to the decay time of $u$ for $u$ tending to zero.
For small $u$, the waiting time distribution of $\tau$ is given by
\begin{equation}
P(\tau)=(N-1)\gamma\exp[-\gamma(N-1) \tau].
\label{eq:waiting_time}
\end{equation}
Averaging the evolution of $u$ over this waiting time in the limit for small $u$ and to leading order in $1/N$ leads to $u\rightarrow [1-1/(8 N^2)] u$. The average number of jumps per unit time is $\gamma N$, so that on average 
\begin{equation}
\frac{d u}{dt}= -\frac{\gamma}{8N} u.
\end{equation}

The same calculation can be repeated including the gain due to the reservoir $r$. The change in $u$ for a single quantum jump, either loss or gain, is still given by Eq.~\eq{total} in the large $N$ limit. The number of jumps however doubles (on average the gain $r$ should compensate for the losses $\gamma$), so that we obtain the Schawlow-–Townes expression for the decay of the coherence:
\begin{equation}
\gamma_c = - \frac{\gamma}{4N}.
\label{eq:st}
\end{equation}
The equal contribution of losses and gain to the laser linewidth was noted by Scully and Lamb~\cite{scully-lamb}.

The equation~\eq{st} is entirely general and does not make any assumptions on the nature of the gain medium. In the present formulation, the robustness of the Schawlow–Townes linewidth comes from the fact that the decoherence per quantum jump does not depend on $\gamma$ or $r$. It is therefore insensitive to the fluctuations in the gain and the strength of gain saturation. 

The physics becomes more complicated when interactions between the reservoir and polaritons are taken into account. This is important in semiconductor lasers where the carrier concentration affects the refractive index and thus shifts the polariton energy. Fluctuations in the reservoir occupancy then affect the laser line width, leading to the well known Henry linewidth enhancement factor~\cite{henry}. We will now proceed to discuss the enhancement of the linewidth due to such interactions. 
They result in an additional phase shift of $u$ under the deterministic evolution, that is taken to be of the form $\Delta \phi = g_R \Delta t$.

\begin{figure}
	\includegraphics[width=0.5\columnwidth]{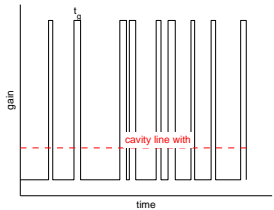}
	\caption{Gain as a function of time in the case of a strong backaction of the cavity dynamics on the reservoir population.}
	\label{fig:gaint}
\end{figure}

A situation in which the linewidth enhancement can be easily computed is the one illustrated in Fig.~\ref{fig:gaint}. Here, the gain due to a single particle in the reservoir is assumed to be much larger than the cavity line width. As a consequence, the reservoir occupation has a very small probability to be occupied by more than one particle. 
The evolution of $u$ under a single gain peak is given by 
\begin{equation}
u\rightarrow \frac{\sqrt{n} e^{-i g_R \tau}u}{\sqrt{n u^2 +(N-1)(1-u^2) \exp[\gamma \tau]}}
\label{eq:total_interactions}
\end{equation}
Averaging again over the waiting time distribution for $\tau$, and adding the decoherence due to the losses, we obtain in the large $N$ limit
\begin{equation}
\frac{d}{dt}u =- \gamma \left[i \frac{g_R}{r} +
\frac{ 1+4 (g_R/r)^2}{4N} 
\right]
 \; u,
\label{eq:henry_large_gainsat}
\end{equation}
that gives a polariton line width
\begin{equation}
\gamma_c = 
\frac{ 1+4 (g_R/r)^2}{4N} \gamma
\label{eq:henry_lw1}
\end{equation}
The $1/N$ scaling with the number of polaritons is of the Shawlow-Townes form \eq{st} and the increase of the linewidth is the Henry linewidth enhancement~\cite{henry}.
The standard deviation of this factor is in a semiclassical approximation, that assumes small fluctuations in the gain medium density~\cite{haug_book} and is limited to large polariton occupation numbers. It thus turns out that the Henry linewidth enhancement is also a robust phenomenon that is insensitive to the details of the system-reservoir coupling.

In our formalism, this semiclassical approximation corresponds to approximating the quantum jump model by a diffusion model~\cite{gardiner_handbook}. The Langevin equations for the photon and reservoir occupations $N_R$ are taken to be
\begin{eqnarray}
d N &=& [ r N_R (N+1) -\gamma N ] dt + \sqrt{r N_R(N+1)} dW_G \notag \\
&& \vspace{1cm} + \sqrt{\gamma N} dW_L, \\
d N_R &=& - r N_R(N+1) dt -  \sqrt{r N_R(N+1)} dW_G.
\label{eq:langevin}
\end{eqnarray}
Here $dW_{G,L}$ are independent Gaussian noise terms with variance $\langle d W_{G,L} dW_{G,L}=dt \rangle$. The stochastic terms represent the shot noise fluctuations of the photon gain and losses. 

In the large photon limit, the Langevin equation~\eq{langevin} can be solved approximately by linearizing around the mean field solution $(N^{(0)},N_R^{(0)})$ : $N=N^{(0)}+\delta N$ and $N_R=N_R^{(0)}+\delta N_R$, yielding
\begin{multline}
\frac{d}{dt}\left( 
\begin{array}{c}
\delta N \\ 
\delta N_{R}%
\end{array}%
\right) =\left( 
\begin{array}{cc}
0 & rN_{0} \\ 
-\gamma  & -rN_{0}%
\end{array}%
\right) \left( 
\begin{array}{c}
\delta N \\ 
\delta N_{R}%
\end{array}%
\right) \\
 +\left( 
\begin{array}{c}
1 \\ 
-1%
\end{array}%
\right) \sqrt{rN_{R,0}N_{0}}dW_{G}+\left( 
\begin{array}{c}
1 \\ 
0%
\end{array}%
\right) \sqrt{\gamma N_{0}}dW_{L}.
\label{eq:Lin}
\end{multline}

The interaction with the reservoir induce a phase shift between the two components of in Eq. \eqref{eq:dtheta}, that is equal to
\begin{equation}
\phi(t)= g_R \int_0^t \delta N_R(t') dt'.
\end{equation}
This phase shift causes a decoherence that is to be added to the Shawlow-Townes decoherence mechanism that leads to a decrease of the absolute value of the first component of the vector Eq. \eqref{eq:dtheta}.

Using the expression for Gaussian variables $\langle e^{i[\phi(t)-\phi(0)]} \rangle = e^{- \langle [\phi(t)-\phi(0)]^2 \rangle /2}$, we can write the first order coherence as
\begin{equation}
g_{\rm res-pol}^{(1)}(t)= \exp\left[-X(t)
\right],
\label{eq:g1}
\end{equation}
where
\begin{equation}
X(t)=\frac{g_R^2}{2} 
\left \langle
\left[
\int_0^t\delta N_R(t')dt'
\right]^2
\right \rangle.
\label{eq:defX}
\end{equation}
Here, the coupling constant $g_R$ quantifies the strength of interactions between reservoir excitons and polaritons. The subcript in Eq. \eqref{eq:g1} indicates that we only consider the effect of the reservoir-polariton interactions on the polariton linewidth. Other contributions to the decoherence, such as the Shawlow-Townes mechanims and the effect of polariton-polariton interactions (see below) have to be added to this decoherence.

The expectation value in the above expression can be rewritten as \cite{anderson_kubo}
\begin{equation}
\left \langle
\left[
\int_0^t\delta N_R(t')dt'
\right]^2
\right \rangle
= 2 \int_0^t dt' (t-t') \Phi_R(t'),
\label{eq:kubo}
\end{equation}
where we have used that the reservoir correlation function 
\begin{equation}
\Phi_R(t_1-t_2) = \langle  
\delta N_R(t_1) \delta N_R(t_2)
\rangle,
\end{equation}
only depends on the time difference.

Eq. \eq{kubo} is a relation that is used in used in the standard Kubo theory for the line width of an emitter with time-dependent frequency \citep{anderson_kubo,kubo}. Two limiting cases of this expression are of particular interest. For times that are short compared to the correlation time of the reservoir fluctuations $\tau_R$, the coherence decreases as a Gaussian
\begin{equation}
X(t) \sim \Phi_R(0) t^2,  
\end{equation}
where for times much larger than the reservoir fluctuation time
\begin{equation}
X(t) \sim  \Phi_R(0) \; \tau_R t \ll  \Phi_R(0)\;  t^2.
\end{equation}
The above inequality is physically described as motional narrowing: for times much longer than the fluctuation time of the reservoir, the decoherence effect on the system is suppressed.

The reservoir correlation function can be easily computed in the Fourrier domain, using the relation
\begin{equation}
\Phi_R(t)=\int \frac{d\omega}{2\pi} e^{-i\omega t} \langle  |\delta N_R(\omega)|^2 \rangle.
\label{eq:phi_R}
\end{equation}
From the Fourrier transform of Eq. \eq{Lin}, one obtains
\begin{equation}
 \langle  |\delta N_R(\omega)|^2 \rangle = \gamma N_0
  \left(
 \frac{\omega^2+2/\gamma^2}{(\omega^2+\Gamma_1^2)(\omega^2+\Gamma_2^2)}
 \right),
 \label{eq:NR_w}
\end{equation}
where the linear damping rates are given by
\begin{equation}
\Gamma_{1,2}=\frac{1}{2}\left(
r N_0 \mp \sqrt{(rN_0)^2-4 \gamma r N_0}.
\right).
\end{equation}

Using Eqns. \eq{g1}, \eq{phi_R} and \eq{NR_w}, the motional narrowing dominated long time behavior of the decoherence induced by the interactions between reservoir particles and the system polaritons is given by 
\begin{equation}
g_{\rm res-pol}^{(1)}(t) \sim \exp\left[
- \left( \frac{g_R}{r} \right)^2 \frac{\gamma}{N_0} T
\right]  \hspace{1 cm} {\rm for } \;\; t\gg  1/\Gamma_1,
\label{eq:g1_nar_res}
\end{equation}
which is despite the very different approximations, identitical to the interaction contribution in Eq. \eq{henry_large_gainsat}.

The Shawlow-Townes contribution to the coherence decay \eqref{eq:st} is to be added to this expression. Mathematically, it comes from the decrease in magnitude of $u(t)$ in Eq. \eqref{eq:dtheta}, where the decay \eqref{eq:g1_nar_res} is due to the scrambling of its phase.

The Gaussian short time behavior is governed by
\begin{equation}
g_{\rm res-pol}^{(1)}(t)= \exp \left(
- \frac{g_R^2}{r^2} \frac{\gamma(2\gamma + r N_0)}{4  N_0} T^2
\right) 
 \hspace{0.5 cm} {\rm for } \;\; T\ll 1/\Gamma_{1,2},
 \label{eq:res_gauss}
\end{equation}
The Gaussian early time decay was not obtained in the case of the large gain saturation, described by Eq.~\eq{henry_large_gainsat}. This is consistent with the fact that the reservoir correlation time vanishes in this approximation.

Entirely analogously, the line broadening due to the interactions between the polaritons can be computed. Within the linearized model \eq{Lin}, the photon fluctuations are
\begin{equation}
\langle |\delta N(\omega) |^2 \rangle = \frac{2\omega^2+(r N_0)^2}{(\omega^2+\Gamma_1^2)(\omega^2+\Gamma_2^2)}
\end{equation}

The contribution to the long-time decoherence is 
\begin{equation}
g^{(1)}_{\rm pol-pol}(t)=\exp\left(- g_p^2
\frac{N_0}{2 \gamma} T
\right) \hspace{0.5 cm}  {\rm for } \;\; T\gg  1/\Gamma_{1},
\label{eq:tau3}
\end{equation}
where $g_p$ is the interaction strength between polaritons.
The dependence on particle number $N_0$ and linewidth $\gamma$ is the same as the one obtained by Eastham and Whittaker \citep{eastham}. 
Note that it scales very differently than the linewidth induced by the interactions with the reservoir \eq{g1_nar_res}. In particular, the line-broadening due to the polariton-reservoir interactions decreases with increasing polariton number, where the broadening due to the polariton-polariton interactions increases with increasing polariton number.

For short times, the coherence decays as
\begin{equation}
g^{(1)}_{\rm pol-pol}(t)= \exp\left(
-g_p^2\frac{(2\gamma+rN_0)}{4 r } T^2
\right)  \hspace{0.5 cm} {\rm for } \;\; T\ll 1/\Gamma_{1,2}.
\label{eq:tau4}
\end{equation}
Compared to the result for reservoir-polariton interactions \eq{res_gauss}, a large polariton population is less favorable for good temporal coherence. 

It is also interesting to notice the different dependence of the coherence on the gain saturation parameter $r$. The coherence times due to the reservoir-polariton interactions in Eqns. \eqref{eq:g1_nar_res}, \eqref{eq:res_gauss} scale as $\sim r^2$. The coherence time in Eq. \eqref{eq:tau4} scales weaker, as $\sim r$, and at long times, the coherence decay \eqref{eq:g1_nar_res} even becomes independent of the gain saturation.

\section{Conclusions}

We have constructed a quantum jump model that is able to describe the dynamics of a coupled quantum-classical system, where the quantum system enjoys gain from the classical one. Our model makes it possible to describe arbitrarily complex dynamics in the reservoir and energy dependent gain. 
As a first application of the theory, we have investigated the linewidth of a single mode polariton condensate. The Shawlow-Townes line width, line broadening due to polariton-reservoir interactions and polariton-polariton intereactions can be described with this formalism. We have shown that the these two mechanisms for line broadening exhibit a different dependence on the number of polaritons and on the reservoir gain saturation.

Our approach can be staightforwardly combined with a full Boltzmann dynamics of the reservoir, so to give an ab initio quantum description of polariton condensation in semiconductor micro and nanocavities. Moreover, the formalism can be applied to spatially extended systems as well, that require to take into account the energy dependence of the gain process \cite{wigner}.

\section{Acknowledgements}
This work was supported by the UA-LP and FWO-Odysseus programs.

\end{document}